# Disk-Galaxy Density Distribution from Orbital Speeds using Newton's Law

Kenneth F. Nicholson  (Caltech Alumni)
email: knchlsn@alumni.caltech.edu

**Abstract**

Given the dimensions (including thickness) of an axisymmetric galaxy, Newton's Law is used in integral form to find the density distributions required to match a wide range of orbital speed profiles. Newton's law is not modified and no dark matter halos are required.  The speed distributions can have extreme shapes if they are reasonably smooth.  Several examples are given.

## 1.  Introduction

Most methods to find the mass distributions causing observed galaxy speed profiles have assumed that the mass inside a given radius can be measured by the speed at that radius, following the Kepler formula $m = v^2 r / G$, identical with that for spheres.  Failures with these methods have brought on attempts to force speed profile matches by adding spherical or elliptical "dark matter" halos that extend well outside the observed envelopes of the galaxies (Sacket, 1999), or by modifying the value of G with distance in a way to keep all matter inside those envelopes (van den Bosch and Dalcanton, 1999) . These methods assume the matter outside a given radius has no gravitational effects on that inside. However for any galaxy the orbital speed at a given radius can be found only by applying Newton's law over the complete volume. The effects of axisymmetric matter outside a given radius on that inside are not negligible for disks as they are for spheres.

Sofue (1999) has computed the surface mass density (SMD) for the Large Magellanic Cloud using both the spherical equivalent method and Poisson's equation. His results show both methods have the same general form for the mass distribution profile and agree in magnitude within a factor of 1.5 to 2. This agreement suggests Poisson's equation also assumes that mass outside a radius has no effect on that inside. These results will be compared later to the method developed here. Binney and Tremaine (1987) show some methods for finding mass distribution from orbital speed profiles, but these use mathematical procedures that have computing difficulties, such as division by zero, difficult table lookups, and series solutions for elliptic integrals or Bessel functions. It is difficult to check the accuracy of any of these methods. To check results the mass distributions obtained must be used to compute the speed profiles used as inputs, using a proven method. To handle a wide variety of galaxy shapes and speed profiles, such a proven method must include a computer program that integrates the effects of all particles in a galaxy, on a test mass at any radius, using Newton's law. To prove the method and program, results must be compared with known theoretical solutions.

## 2.  Program Development

Given galaxy dimensions of thickness vs. radius, the forward problem is defined here as finding the orbital speed profile for a given density distribution, and the reverse problem as finding the density distribution for a given speed profile.  For these problems it is assumed that almost all the galaxy mass is contained within the defined dimensions, and relativistic effects are negligible.

As compared with other methods for this problem, such as those in Binney and Tremaine (1987), many concepts are changed.

Since SMD is not a realistic property of matter in a galaxy, the concept of a very thin disk was discarded in favor of solutions for density, even if the galaxy thickness has to be estimated. Density can be used to find the average distance between sun-sized stars and should correlate better with star formation than SMD. In the neighborhood of our own sun, density results can be checked against other estimates. At the center of the galaxies the distance between stars allows a reasonable guess about the presence of a black hole. Also curves of mass vs. radius are much better than log (SMD) vs. radius for characterizing the gravitational effects of a galaxy, and can be related in a general way to the speed profile

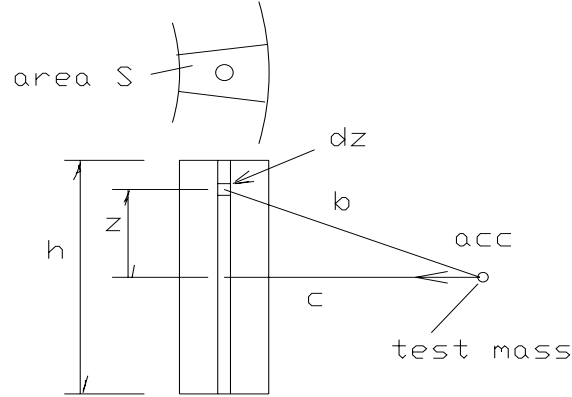

Figure 1. Fundamental segment

The fundamental segment used is shown in figure 1, with all mass concentrated as a rod located halfway between ring edges. The rod should be at the centroid of the section area, but it eases computing and causes little error where it is. The gravitational effects of the segment on the test mass are analytic, saving much computing.

$$acc = G*rho*\int_{-h/2}^{h/2} \frac{(S\,dz)\,c}{b^3} = \frac{G\,ddm}{c\,\sqrt{c^2+(h/2)^2}} \quad , \text{ where } ddm = rho*S*h \qquad (1)$$

In previous methods (Binney and Tremaine, 1987) the force acting between a volume increment and a test mass becomes infinite as the distance between becomes zero, and it does here also. This "division-by-zero problem" has led to much grief in the past Some procedures simply delete that point, but that still leaves the high forces as the point is approached. A simple thought experiment shows that in reality the force does not become infinite. It goes through zero as the test mass passes through an isolated real segment made up of diffuse particles. For a good approximation to deal with this, the test mass here is made to pass equidistant between two segments at all times, causing the force from these two to go to zero at the minimum distance.

For the reverse problem, the methods are again different. Given dimensions including thickness, a first estimate of the average density is used for all galaxy radii to start, and the orbital speeds computed at the outer edges of each ring. These are compared with the measured speeds, and the corresponding densities corrected by an amount set by problem stability. A density change in one ring changes the speeds for all other rings, but the procedure converges well. This process continues until the computed and measured speeds all agree within a very small error limit. The error limit is chosen to make the density errors all very small fractions of the maximum density. The density distribution is then close to the correct distribution to cause the input speed profile. It is a feedback scheme applied to the densities of each ring (from 20 to 50), thus solving a difficult nonlinear integral equation. With a 450 mhz desktop computer this turns out to be fast and easy, with excellent accuracy.

Results are computed for density and mass distributions, plus total mass for the reverse problem. For the forward problem the speed profile is computed, along with normalizing constants to generate dimensionless results. These dimensionless results are used for studies to show overall trends, because the results are then general and can be applied to galaxies of any size. This allows the conclusion that there could be many small dark galaxies.

For gravitational purposes it makes no difference what the material is. It could be sand. In particular there is no requirement for stars or light, or any correlation between mass and luminosity.

### 3. Notation and equations

No subscripts or Greek letters are used for computing or plots, so most of the variables have more than one letter. With an ending of d (always lower case), the variable is dimensionless, and for those the normalizing constants are: rmax for all dimensions, the rim Kepler speed (vkr) for all speeds, rim Kepler acceleration (Akr) for acceleration, total mass (mtot) for mass, and average density (rhoav) for density. Surface mass density (SMD) is an exception and is always capitalized.

```
A      = acceleration of test mass,  plus toward galaxy center, pc/yr^2
Akr    = Kepler accelertion at the rim, G*mtot / rmax^2, pc/yr^2
ddm    = mass in a single segment, rho*h*r*dth*dr, msuns
dr     = radial thickness of a ring, pc
dth    = pi / 180 for all cases, rad
G      = gravitational constant,  4.498E(-15)  pc^3/(msuns*yr^2)
h      =  galaxy thickness at radius r,  pc
m      = galaxy mass inside rm,  msuns
mtot   = galaxy total mass,  msuns
msun   = mass of our sun,  1.989E33 gms
pc     = parsec, 3.08568E13  kms, 3.26151 light years
pi     = 3.14159
pytks  = multiplier to change v(pcs/yr) to v(kms/sec),  9.7778E5  km/pc*yr/sec
r      =  radius to centerline circle of a ring , r = rm-dr/2, pcs
rho    = density of a ring,  msuns/pc^3
rhoav  =  average density of galaxy,  mtot / voltot, msuns/pc^3
rm     = radius to outer edge of a ring, pcs
rmax   = galaxy maximum radius, pcs
rt     = radius to test mass from galaxy center, pcs
            the test mass is located at the outer edge of each ring when on the galaxy
SMD    = surface mass density, msuns / pc^2
th     =  angle from galaxy radial line of test mass to radial line of ring segment, rad
v      = sqr(A*rt) * pytks, computed orbital speed at test mass radius rt,  kms/sec
vkr    = sqr(Akr * rmax) * pytks,  Keper orbital speed at the rim, kms/sec
vm     = measured speed, km/sec
vmax   = maximum measured speed, km/sec
voltot = galaxy total volume, pc^3
```

For the forward problem, given dimensions and the density distribution, using the concepts above and symmetry, orbital speed is found for each test mass location by integrating the effects around each ring, then summing the effects of the rings.

$$v^2 = (pytks)^2 \bullet rt \bullet \int_0^{r\,max} 2 \int_0^{pi} \frac{G\ ddm}{c\ \sqrt{c^2 + (h/2)^2}} \bullet \frac{(rt - r\cos(th))}{c} \qquad (2)$$

$$= (pytks)^2 \bullet rt \bullet \sum_1^{Nr} 2 \sum_1^{180} \frac{G\ ddm}{\sqrt{c^2 + (h/2)^2}} \bullet \frac{(rt - r\cos(th))}{c^2} \qquad (3)$$

where   $c^2 = (r\sin(th))^2 + (rt - r\cos(th))^2$

Nr  =  number of rings, 20 to 40 for the reverse problem, up to 100 for forward

For the reverse problem , given dimensions and measured speeds, the same equation is used to compute the speeds for each ring , and density there is corrected for the next cycle.

errv  = (vm-v) / vmax ,  f = 0.75 * errv ,

limit abs(f) < 0.5 ,  rho(i) = (1+f)*rho(i-1) for each cycle  i

## 4.  Proof of equations and code for the forward problem

Proof is done by comparing results with a sphere and with an isolated ring.  Also speeds for several galaxy shapes are computed to check trends and build confidence. The shapes considered are described below and in Figure 2. They include the Milky Way and the Large Magellanic Cloud (LMC) for convenience.

rmax:      17000 for Milky Way,  8000 for LMC,  10000 for all others, pcs

sphere:    hd = 2 * sqr (1- rd^2) , rho = constant

ring:      hd  = 0.05 ,  solid for rd = 0.99 to 1.00 , rho = constant flat disk:  hd = 0.05 ,   rho = constant thick:     rho = k * h ,  hd = 0.2*sqr(1 - (rd/0.15)*2) ,  for rd < 0.075 ,
           then    hd = 0.05 + 0.1232 * exp ( - 3.486 * (rd - 0.075))

thinner:    rho = k * h,  hd = 0.2 * sqr (1 - (rd/0.15)^2) ,  for rd < 0.075 ,
            then     hd = 0.025 +  0.1482 * exp ( - 4.413 * (rd - 0.075))

Milky Way:  rho to be found ,  hd = 0.1288*sqr(1 - (rd/0.0644)^2) ,  for rd < 0.0322 ,
              then     hd = 0.11158*exp ( - 2.315*(rd - 0.0322)) , for rd < 0.2857 ,
              then     hd = 0.0620 -  0.0727 * (rd - 0.2857)

LMC:  rho to be found,  assume same shape as Milky Way

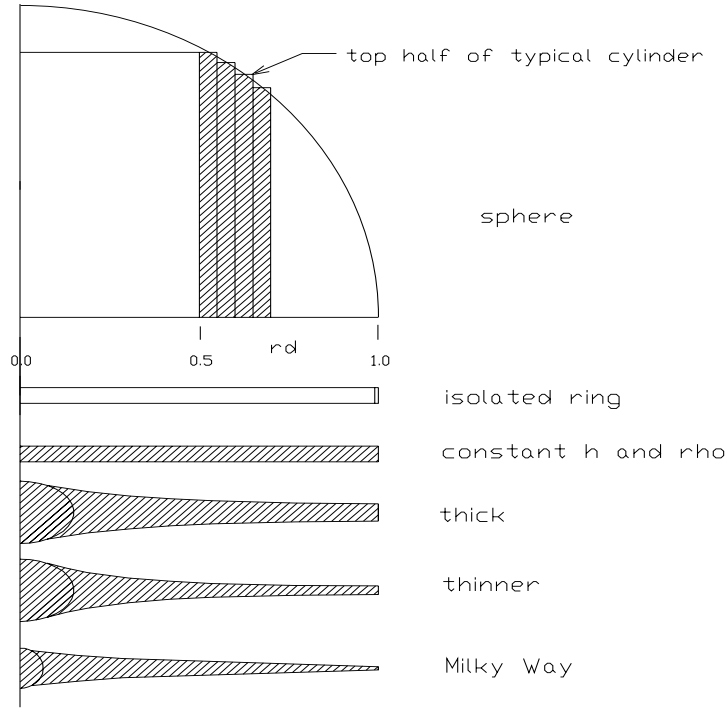

Figure 2. Galaxy shapes

The spherical check (figure 3) is close, although the volume is slightly high and rim velocity slightly low. Near the rim the cylinder volumes are too large as compared to the actual spherical portions. The errors cannot be seen on the plot, and this error source is gone for disks.

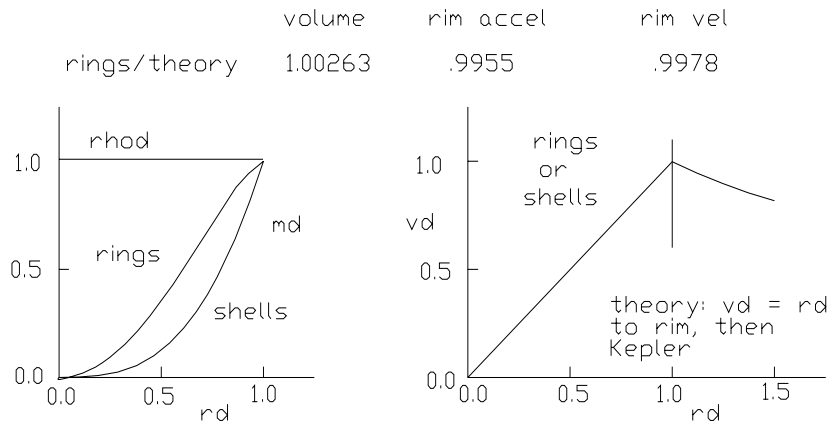

| | volume | rim accel | rim vel |
|---|---|---|---|
| rings/theory | 1.00263 | .9955 | .9978 |

Figure 3. Sphere using 100 rings

Proof is quite good for radii inside and outside the single ring (figures 4.1 and 4.2), as far as elliptic integrals can be used. Right at the ring (rd = 0.995) the Kepler acceleration results and the answer looks correct, but I have not yet proved it. The elliptic integral there is infinite. Using a planetary circle of spheres and increasing the number, the acceleration converges toward the Kepler value then gradually goes beyond toward infinity, like the elliptic integrals. The difference appears to be the use of the approximation for a ring using thickness and diffuse segments , instead of treating each segment like a sphere. This is the first solution I have seen for the acceleration at an isolated ring, and until shown to be wrong it must serve as a second proof.

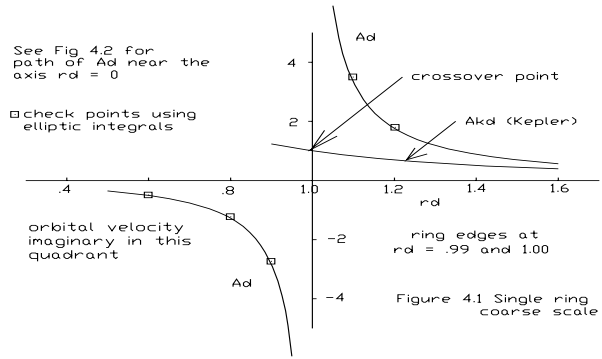

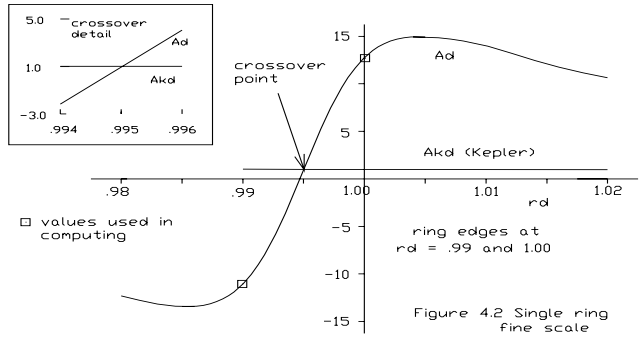

For elliptic integrals the dimensionless acceleration for the check points is reduced to:

$$Ad = \frac{1}{pi * rd}\left[\frac{K(k)}{1+rd} - \frac{E(k)}{1-rd}\right] , \text{ where } k^2 = \frac{4 * rd}{(1+rd)^2} \tag{4}$$

## 5. Results and trends with the forward problem

Figure 5 shows that the speeds for a flat disk get much larger than predicted using spherical shells. Using spherical shells, speeds are set by the mass inside a radius, becoming equal to the Kepler value at the rim. So using spherical shells, measured speeds for a disk like this would result in a computed mass profile too low at the low radii and too high near the rim. Total mass would be high by a factor of nearly 3. This basic flat-disk result also appears to be a new solution. It can not be found by the methods in Binney and Tremaine.

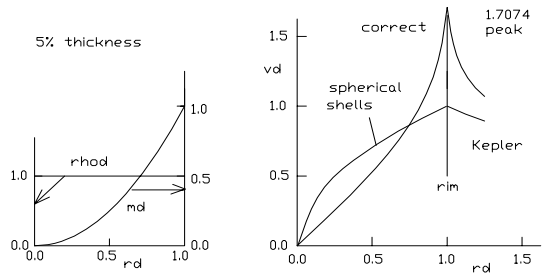

Figure 5. Comparison of methods for a flat disk

Rim speed varies significantly with thickness for this type disk (figure 6), showing another reason why thickness should be included in these studies.

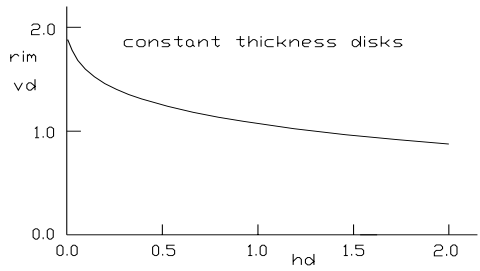

Figure 6. Effects of Thickness on Rim Speed

To show trends as mass is moved toward the center, 3 galaxies are compared in Figure7. Shapes are shown in figure 2. The speed profiles shift toward the Kepler shape for a concentrated mass at the center and all cases become asymptotic to the Kepler values beyond the rim as they should. Relative densities are proportional to thickness and increase toward the center because of shape only. The abrupt speed increase near the rim is related to rim thickness, and the effect decreases from thick to thinner.

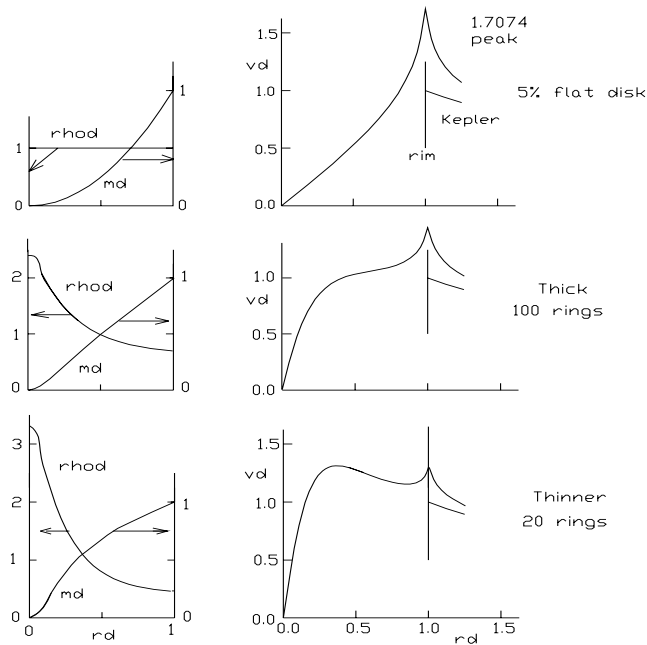

Figure 7. Comparison of 3 galaxies

In this dimensionless form of figure 7 it is noticed that the shape of the speed curves correlate fairly well with the shape of the mass curves. Describing the mass curve as radius to the power x , the correlation would be something like this:

flat disk           x = 2.0 ,      vd ~ rd

thick              x = 1.0 ,      vd ~ slightly  increasing to rim

$$md = (rd)^x$$

thinner            x = 0.7 ,      vd ~ near Kepler

all mass at center    x = 0.0 ,      vd ~ Kepler

The thinner example has only 20 rings because it is used also to provide proof for the reverse problem. It also shows that only 20 rings are needed for smooth results.

## 5.  Proof of equations and code for the reverse  problem

The speeds computed for the thinner galaxy and its dimensions were used as input, with the output being density vs. radius, to try to match the original density distribution.  By matching speeds within vmax * 1E-5, the densities were returned with errors of less than  rhomax * 1E-4. Errors in total mass were less than mtot * 1.5E-6. These results are considered proof enough.

## 6.  Milky Way

Input data are from a survey article by Bok (1981). Dimensions (figure 2) are from measurements of a computer-generated side view in that article, done by Babcall and Soneira at the Institute for Advanced Study.  The measured speeds are in figure 8, and the downward sloping line past 10 kpcs is the Kepler result for the mass inside that radius.

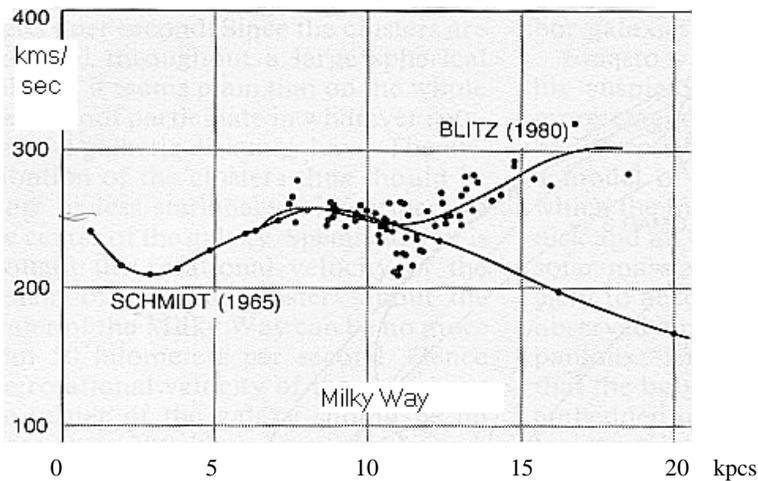

Figure 8. Milky Way measured speeds

Speed inputs were faired from the measured data and rounded to 3 significant figures. Twenty rings are used and  rmax chosen at 17000 pcs. Matching speeds are shown to 3 decimal places to see the errors. The results are listed here and plotted in figure 9.  SMD is included for comparisons with other results. There may be more recent results for orbital speeds, but these will serve well to demonstrate the method.

| ring | r, pcs | measured vm, kms/sec | matched v, kms/sec | rho msuns/pc^3 | SMD msuns/pc^2 | |
|------|--------|----------------------|--------------------|-----------------|-----------------|---|
| 1 | 850 | 240.000 | 240.000 | 4.152 | 8377.9 | |
| 2 | 1700 | 222.000 | 220.000 | 0.619 | 1064.2 | |
| 3 | 2550 | 209.000 | 209.000 | 0.477 | 729.3 | |
| 4 | 3400 | 210.000 | 210.000 | 0.467 | 637.2 | |
| 5 | 4250 | 220.000 | 220.000 | 0.466 | 566.3 | |
| 6 | 5100 | 231.000 | 231.000 | 0.449 | 485.9 | |
| 7 | 5950 | 242.000 | 242.001 | 0.421 | 423.0 | |
| 8 | 6800 | 253.000 | 253.001 | 0.387 | 365.1 | |
| 9 | 7650 | 260.000 | 260.001 | 0.342 | 301.4 | |
| 10 | 8500 | 257.000 | 257.001 | 0.287 | 235.6 | approx sun location |
| 11 | 9350 | 250.000 | 250.002 | 0.258 | 195.5 | |

| 12 | 10200 | 245.000 | 245.003 | 0.256 | 178.3 |
| 13 | 11050 | 248.000 | 248.003 | 0.276 | 175.2 |
| 14 | 11900 | 254.000 | 254.002 | 0.292 | 167.4 |
| 15 | 12750 | 261.000 | 261.000 | 0.306 | 156.5 |
| 16 | 13600 | 271.000 | 270.999 | 0.324 | 145.7 |
| 17 | 14450 | 280.000 | 279.999 | 0.333 | 128.9 |
| 18 | 15300 | 289.000 | 288.999 | 0.339 | 110.5 |
| 19 | 16150 | 300.000 | 299.999 | 0.341 | 90.1 |
| 20 | 17000 | 310.000 | 309.999 | 0.295 | 59.6 |

At the sun radius the density is found to be 0.287 msuns per cubic parsec, compared with 0.18 in Binney and Tremaine Table 1-1. Yet if the stars make up 24.4% of the mass as in that table , the 0.287 density translates to about 7.91 light years average distance between sun-size stars. This seems about right.

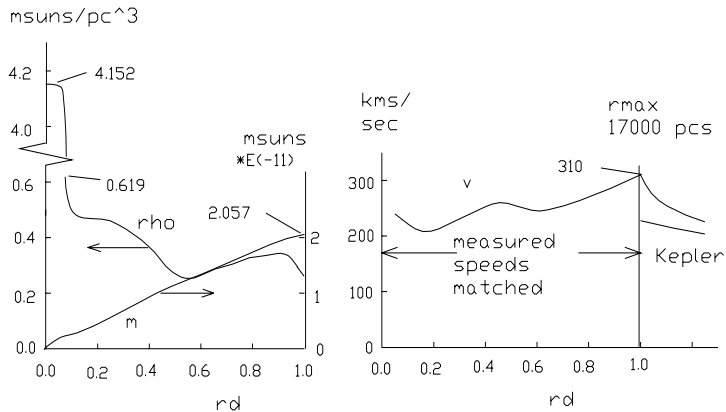

Figure 9. Milky Way

The plotted results in figure 8 show a galaxy total mass of 2.057E11 sun masses, much lower than values usually quoted , which often exceed 3E11when dark-matter halos are used to match orbital speeds. At the center of the Milky Way, the density is quite high, 4.152 msuns per cubic parsec. Assuming stars make up 75% of the mass, the average distance between sun-size stars becomes about 2.23 light years. This is still plenty of clearance between stars and suggests that the Milky Way center is a typical elliptical collection of stars rather than a black hole.

## 8. Large Magellenic Cloud

Figure 9 compares the methods of spheres, Poisson's equation, and that developed here (using Newton's inverse square law, NISL). The orbital speeds and SMD data were read from the plots of Yoshiaki Sofue's paper. To get dimensions, his value of maximum radius, and the Milky Way shape were used. So the thickness is estimated to get the density, but the mass distribution using SMD does not depend on thickness, and the dependence using NISL is very weak.

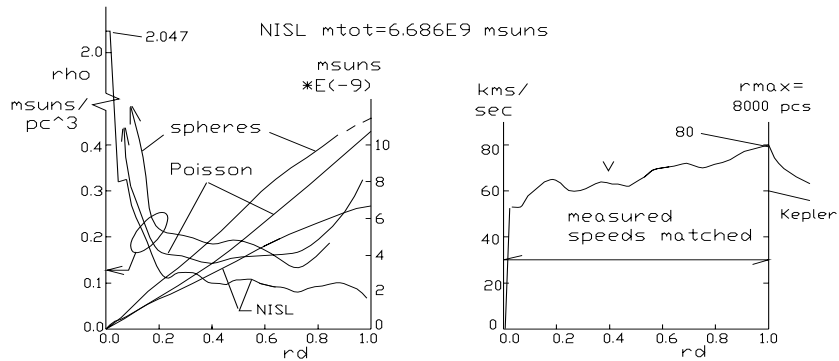

Figure 10. Comparison of methods for the Large Magellanic Cloud

The figure shows trends similar to the Milky Way for density using NISL, but higher values and strongly increasing trends near the rim for both of the other two methods. At the center the NISL method shows a density about half that of the Milky Way, while the others continue on up to larger values. Total masses of the other methods exceed that of NISL by about 60%.

From these results it is seen that the use of Poisson's equation and spheres lead to similar answers, and both appear to neglect the effects of material outside a radius on that inside. I believe that Poisson's equation is being misapplied for this purpose. There is no doubt about the use of spheres. They don't apply here.

The total mass obtained, 6.686E9 msuns, implies far too much accuracy for an overall answer, but it is that accurate for an ideal set of the dimensions and speeds supplied. The same comments apply to the Milky Way results.

## 7.  Conclusions

Newton's law needs no correction, and dark-matter halos are not needed, to compute galaxy density distributions from orbital speeds. Using measured speeds and defined dimensions, the results for the density and mass distributions using the methods shown here will be as good as the inputs.

The use of spherical shells or Poisson's equation are not suitable for this purpose.

There is no mystery about orbital speeds being constant to large radii, or rising near galaxy rims.

## References

Binney, J. and Tremaine, S. 1987, Galactic Dynamics (book, Princeton University Press)
Bok, R.J., The Milky Way Galaxy,  Scientific American, March 1981
Corbelli, E. and Salucci, P., astro-ph / 9909252
Dalcanton, J. J. and Bernstein, R. A., astro-ph / 9910219
Sacket, P. D., astro-ph / 9903420
Sofue, Y., astro-ph / 9906222
van den Bosch and Dalconton , astro-ph / 9912004